\documentstyle[preprint,eqsecnum,prd,aps,epsf]{revtex}
\begin{document}

\def\grtsim{\,\,\rlap{\raise 3pt\hbox{$>$}}{\lower 3pt\hbox{$\sim$}}\,\,}
\def\lsim{\,\,\rlap{\raise 3pt\hbox{$<$}}{\lower 3pt\hbox{$\sim$}}\,\,}

\draft
 
{\tighten
\preprint{\vbox{\hbox{CALT-68-2084}
                \hbox{hep-ph/9611263} 
		\hbox{\footnotesize DOE RESEARCH AND}
		\hbox{\footnotesize DEVELOPMENT REPORT} }}
 
\title{Non-linear Sigma Model Solutions for the Disoriented Chiral Condensate at ${\cal O}(p^4)$
}
 
\author{Hooman Davoudiasl\footnote{E-mail address: hooman@theory.caltech.edu}}
 
\address{California Institute of Technology, Pasadena, CA 91125 USA}

\maketitle 

\begin{abstract}
Boost-invariant (1+1) dimensional solutions for the Disoriented Chiral Condensate (DCC) are obtained numerically, in the context of the $SU(2)_L \otimes SU(2)_R$ non-linear sigma model at ${\cal O}(p^4)$ in the momentum expansion.  We ignore the mass terms in the Lagrangian, as we are mainly interested in the behavior of the solutions for small values of proper time $\tau$.  The solutions obtained at ${\cal O}(p^4)$ are matched to those of ${\cal O}(p^2)$ at a late proper time $\tau \gg 1/m_\pi$, where $m_\pi = 140$ MeV is the mass of the pion.  We find that at ${\cal O}(p^4)$ the solutions for the DCC do not have singular behavior at early proper times $\tau \ll 1/m_\pi$.  The solutions indicate that for $\tau \lsim (0.5-0.8)$ fm the ${\cal O}(p^4)$ corrections become important.  We take the sizes of the field derivatives to be indicators of the validity of the momentum expansion. Thus, we deduce that the ${\cal O}(p^4)$ solutions can be used to represent the qualitative behavior of the DCC down to proper times of about 0.2 fm.  Since below $\sim 0.2$ fm the formalism is not reliable, we conclude that the inclusion of higher order terms beyond ${\cal O}(p^4)$ is not needed to extend the validity of the solutions to earlier proper times.           
\end{abstract}  

}

\newpage

\pagenumbering{arabic} 

\section{Introduction}

In quantum chromodynamics (QCD), the Lagrangian for strong interactions is invariant under global $SU(N_f)_L \otimes SU(N_f)_R$ chiral transformations, where $N_f$ is the number of massless flavors.  However, the non-zero masses of quarks in the Standard Model explicitly break this global symmetry.  In addition, chiral symmetry is dynamically broken at a scale $\Lambda_\chi
\sim 1$ GeV where the vacuum expectation value of the quark bilinear $\langle \bar q^i q_j \rangle$ becomes non-zero.  Thus, for $N_f$ massless quarks and at energies above $\Lambda_\chi$, the $SU(N_f)_L \otimes SU(N_f)_R$ symmetry of strong interactions is restored.  The restoration of chiral symmetry at non-zero temperatures has been studied in the context of the theory of phase transitions and critical phenomena \cite{PTC}.  According to universality arguments and numerical simulations, in order for the chiral transition to be a second order phase transition, only two quark flavors, namely the up and the down quarks, can be treated as massless \cite{PTC}.  Since the up and the down quark masses are nearly zero compared to the scale of symmetry breaking, the global $SU(2)_L \otimes SU(2)_R$  is an approximate symmetry of QCD at energy scales above $\Lambda_\chi$.

If the masses of the quarks were zero the condensate $\langle \bar q^i q_j \rangle$ would not have any preferred direction in the vacuum, under $SU(2)_L \otimes SU(2)_R$.  However, the small but non-zero quark masses provide the QCD vacuum with a direction.  It has been argued that in certain high energy collisions, such as high energy $\bar p p$ \cite{BKT} or relativistic heavy ion collisions \cite{AAA,AR,BK1,RW}, domains of chiral condensates that do not point in the direction of the QCD vacuum may form and grow to a volume of (a few fermi)$^3$.  These domains are referred to as Disoriented Chiral Condensates (DCC's). The eventual decay of DCC's into a large number of pions is predicted to have a distinct experimental signature, namely large fluctuations in the ratio $f$ of neutral pions to the total number of pions.  

The non-linear sigma model, based on $SU(2)_L \otimes SU(2)_R$, has the essential features necessary for describing low energy QCD phenomena, and has been used to describe the evolution of the DCC after its formation \cite{AAA,AR,BK1,BK2,HS,S}.  In Refs. \cite{BK1,HS}, the authors assume that a relativistic collision can be described by two thin infinite slabs, representing the highly Lorentz-contracted hadrons or nuclei, that collide at the center of mass, and continue along the beam axis in opposite directions. The DCC forms in the region of spacetime between the receding hadrons or nuclei.  These authors find boost-invariant (1+1) dimensional classical pion field solutions of the non-linear sigma model at the leading order in a derivative expansion, without the mass term.  The solutions they obtain exhibit violent oscillations for small values of the proper time $\tau$, which can be taken as 
a sign of the breakdown of the formalism at early proper times, where higher order effects become important.

In this paper, we include the four-derivative terms, ignoring the mass terms, in the Lagrangian for the $SU(2)_L \otimes SU(2)_R$ non-linear sigma model.  We numerically solve the Euler-Lagrange equations to obtain the DCC solutions which are shown to be non-oscillatory at early proper times, $\tau \ll 1/m_\pi$, and we present an analytic explanation for this behavior.  We also show that the corrections from terms with four derivatives are important at proper times earlier than about (0.5-0.8) fm.  The ${\cal O}(p^4)$ solutions for the fields parametrizing the DCC have derivatives that diverge for $\tau \to 0$, as in the case of the previously obtained ${\cal O}(p^2)$ solutions.  However, the proper times at which the magnitudes of the derivatives are deemed too large for a reliable momentum expansion are smaller than those of the leading order solutions.  In the rest of this paper, by ${\cal O}(p^2)$ solutions we mean those obtained from the Lagrangian with terms that have at most two derivatives.  The solutions that are obtained from the Lagrangian that contains terms quadratic and quartic in derivatives are referred to as the ${\cal O}(p^4)$ solutions.  

In the next section, we outline the formalism of the non-linear sigma model at ${\cal O}(p^4)$.  In Section III, we present the ${\cal O}(p^2)$ solutions obtained in Refs. \cite {BK1,HS}.  Section IV, where we present our numerical solutions, includes a discussion of our choice of boundary conditions and a comparison of our solutions with those of the leading order.  Section V contains a summary of our results and some concluding remarks.

\section{The Non-linear Sigma Model at ${\cal O}({\lowercase{p}}^4)$}

In this section, we establish the formalism used in this paper to study the evolution of the DCC at the next to leading order in momentum expansion.  We use notation similar to that of Ref. \cite{HS}.  To represent the effects of chiral symmetry breaking and the pion fields, we introduce the fields $\sigma (x)$ and ${\vec \pi} (x)$, respectively.  We represent the pion field by

\begin{equation}
{\vec \pi} (x) = f \vec n (x) \sin \theta (x),
\label{pi}
\end{equation}
where at leading order $f$ is the pion decay constant $f_\pi = 93$ MeV, $\vec n$ is a unit isovector field $|\vec n|^2 = 1$, and the field $\theta$ is an angle.  The isovector $\vec n$ determines the orientation of the pion field in isospace.   The $\sigma$ and $\vec \pi$ fields are related by 
$\sigma^2 + |\vec \pi|^2= f_\pi^2$.  We define the field $\Sigma$ by
\begin{equation}
f_\pi \Sigma = \sigma + i \vec \tau \cdot \vec \pi,
\label{Sig}
\end{equation}
where $\tau^i$ for $i = 1, 2, 3$ are the Pauli matrices.

The Lagrangian for the non-linear sigma model at ${\cal O}(p^2)$, without the mass terms, can then be written as
\begin{equation}
{\cal L}^{(2)} = {f_\pi^2 \over 4} Tr(\partial_\mu \Sigma \partial^\mu 
\Sigma^\dagger).
\label{L2}
\end{equation}
In terms of the $\theta$ and the $\vec n$ fields, we have
\begin{equation}
{\cal L}^{(2)} = {f_\pi^2 \over 2}(\partial_\mu \theta \partial^\mu \theta +
\sin^2\theta \,\, \partial_\mu \vec n \cdot \partial^\mu \vec n) +  
{\lambda f_\pi^2 \over 2}(|\vec n|^2 - 1),
\label{L2tn}
\end{equation}
where $\lambda$ is a Lagrange multiplier.  

To go to a higher order in the momentum expansion, ${\cal O}(p^4)$, we should include operators with four derivatives, or two derivatives and one insertion of the quark mass matrix.  However, we continue to ignore the mass terms, since we are mainly interested in the early evolution of the DCC, which corresponds to regions of large momenta.  The  ${\cal O}(p^4)$ contribution ${\cal L}^{(4)}$ to the Lagrangian of the system is then given by
\begin{equation}
{\cal L}^{(4)} = \beta_1\left[Tr(\partial_\mu \Sigma \partial^\mu \Sigma^\dagger)\right]^2 + \beta_2 Tr(\partial_\mu \Sigma \partial_\nu \Sigma^\dagger)Tr(\partial^\mu \Sigma \partial^\nu \Sigma^\dagger), \label{L4}
\end{equation}
where, in the notation of Ref. \cite{DGH}, $\beta_1 = \alpha_1 + \alpha_3/2$, $\beta_2 = \alpha_2$, and $\alpha_i, \, \, i = 1, 2, 3$, are the coefficients of the ${\cal O}(p^4)$ terms for the chiral Lagrangian under $SU(3)_L \otimes SU(3)_R$ \cite{DGH}.
In terms of $\theta$ and $\vec n$, we have for ${\cal L}^{(4)}$
\begin{eqnarray}
{\cal L}^{(4)}\!\!&=&\!\!4 \beta_1 \left[\left(\partial_\mu \theta
\partial^\mu \theta\right)^2 + 2\sin^2\theta\left(\partial_\mu \theta \partial^\mu \theta\right) \left(\partial_\nu {\vec n} \cdot \partial^\nu {\vec n}\right) + \sin^4\theta\left(\partial_\mu {\vec n} \cdot \partial^\mu {\vec n}\right)^2 \right]
\nonumber \\
\!\!&+&\!\!4\beta_2\left[\left(\partial_\mu \theta \partial^\mu \theta\right)^2 +\sin^4\theta\left(\partial_\mu {\vec n} \cdot \partial_\nu {\vec n}\right)
\left(\partial^\mu {\vec n} \cdot \partial^\nu {\vec n}\right)
+2\sin^2\theta \, \partial_\mu \theta \partial_\nu \theta
\left(\partial^\mu {\vec n} \cdot \partial^\nu {\vec n}\right)
\right].
\label{L4tn}
\end{eqnarray}

The coefficients $\beta_{1,2}$ get renormalized by one-loop diagrams coming from the ${\cal O}(p^2)$ Lagrangian, including the mass terms, and are phenomenologically determined.  In this work, we are only interested in the classical behavior of the pion field, and hence we do not consider the loop effects.  As our results represent the qualitative behavior of the DCC, we only give an order of magnitude estimate for the typical size $|\beta|$ of $\beta_1$ and $\beta_2$, and ignore the quantum corrections coming from loop diagrams.  We expect this approximation to be qualitatively valid, since by large $N_c$ arguments, where $N_c$ is the number of colors, the effects of loop corrections are suppressed by $N_c^{-1}$.

In order to estimate the $\beta_{1,2}$, we demand, in a systematic momentum expansion of the Lagrangian
\begin{equation}
{\cal L}_p = \left({f_\pi^2 \over 4}\right) p^2 + |\beta| \, p^4 + {\cal O}(p^6), 
\label{Lp}
\end{equation}
that all terms be of the same order at the chiral symmetry breaking scale $\Lambda_\chi$.  That is,
\[
|\beta| \Lambda_\chi^4 \sim {f_\pi^2 \over 4} \Lambda_\chi^2,
\]
and hence
\begin{equation}
|\beta| \sim {f_\pi^2 \over 4\Lambda_\chi^2}.
\label{al}
\end{equation}
For $f_\pi = 93$ MeV and $\Lambda_\chi = 1$ GeV, we get 
\begin{equation}
|\beta| \sim 2 \times 10^{-3}.
\label{alnum}
\end{equation}
For phenomenologically relevant energy scales, $\beta_1 + \beta_2 > 0$ \cite{DGH}.  Since in this paper it is the combination $\beta_1 + \beta_2$ that appears in the equations of motion, we take $\beta=\beta_1 + \beta_2>0$.

\section{The ${\cal O}({\lowercase{p}}^2)$ Solutions}

The ${\cal O}(p^2)$ Euler-Lagrange equations of motion derived from Eq.(\ref{L2tn}) are 
given by 
\begin{equation}     
\Box \theta = \sin \theta \cos \theta \, \partial_\mu \vec n \cdot 
\partial^\mu \vec n 
\label{Op2th}
\end{equation} 
and
\begin{equation}
\partial_\mu(\sin^2 \!\theta \, \partial^\mu \vec n) = \lambda \vec n.
\label{Op2n}
\end{equation}
In Refs. \cite{BK1,HS}, the authors have assumed that the DCC solutions for a relativistic hadronic or nuclear collision have transverse symmetry with respect to the beam direction, where the colliding particles are idealized as two highly Lorentz-contracted slabs of infinite transverse extent.  This idealization makes the problem (1+1) dimensional.  We take these dimensions to be time $t$ and the beam direction $x$.  With the further condition of boost-invariance, the DCC solutions \cite{BK1,HS} become functions of only proper time $\tau = \sqrt{t^2 -x^2}$, that is ${\vec n} = {\vec n}(\tau)$, and $\theta = \theta (\tau)$.  Note that for a function $\phi=\phi (\tau)$ we have $\partial_\mu \phi =( x_\mu/\tau)\phi^{\prime}$, where a prime denotes a derivative with respect to $\tau$; $\phi^{\prime} \equiv d\phi/d\tau$.  Here, we simply mention the solutions to Eqs.(\ref{Op2th}) and (\ref{Op2n}); the details of the solution are found in Refs. \cite{BK1,HS}.  

The angle $\theta$ is given by
\begin{equation}
\cos\theta(\tau) = (b/\kappa)\cos[\kappa \ln (\tau/\tau_0) + \bar \theta_0],
\label{costh}
\end{equation}
where $\kappa \equiv \sqrt{a^2 + b^2}$, $a\equiv |\vec a|$, and $b\equiv |\vec b|$; $\vec a$ and $\vec b$ are two arbitrary constant vectors.  In addition, $\cos \bar \theta_0 = (\kappa / b) \cos\theta(\tau_0)$, and $\tau_0$ is some arbitrary proper time.  The choice of the coordinates in isospace is such that $\vec a$, $\vec c = \vec b \times \vec a$, and $\vec b$ define a right-handed coordinate system, and $n_a = 0$.  We also have
\begin{equation}
n_b(\tau) = ({a \over b})\sqrt {{\kappa^2 \over a^2} - {1 \over \sin^2 \theta(\tau)}} \, . 
\label{nb}
\end{equation}
The solution for $n_c(\tau)$ is obtained from the constraint $|\vec n|^2 = 1$.  In this work, we take $\cos \theta (\tau_0) = b/\kappa$, implying 
$\sin \bar \theta_0 = 0$ \cite{HS}, which will make the above solutions (\ref{costh}) and (\ref{nb}) of the same form as those presented in Ref. \cite{BK1}.  Note that for $a=0$, we have ${\vec n} =$ Constant as the solution.  In the rest of this work, we refer to $n_a$, $n_b$, and $n_c$ by 
$n_1$, $n_2$, and $n_3$, respectively.

The solution for $\theta(\tau)$ given by Eq.(\ref{costh}), oscillates rapidly near $\tau = 0$.  Small values of $\tau$ correspond to regions in spacetime that are close to the highly energetic nuclei.  We expect the theory to break down for small $\tau$, corresponding to large momenta, since our theory is valid only at low momenta.  Thus, it may be tempting to interpret $\tau_0$ as a typical proper time below which the theory becomes unreliable \cite{BK1}, as signaled by the ``rapid'' oscillations.  However, the onset of these ``rapid'' oscillations is scale-dependent, and cannot reliably determine $\tau_0$, in the above sense.  
\section{The ${\cal O}({\lowercase{p}}^4)$ Solutions}

In this section, we will show that the solutions to the ${\cal O}(p^4)$ equations show no divergent behavior near $\tau = 0$.  This does not mean that we can trust the qualitative behavior of the solutions for arbitrarily small  $\tau$.  Instead, we note that our theory is an expansion in derivatives, and thus we will take the magnitudes of the derivatives of the pion field parameters $\theta$ and $\vec n$ to be better indicators of the range of the validity of our solutions.  We will thus demonstrate that the ${\cal O}(p^4)$ solutions stay reliable down to a length scale of about 0.2 fm.   

The Euler-Lagrange equations of motion for $\theta$ and $\vec n$ are derived from the Lagrangian $\cal L$, given by
\begin{equation}
{\cal L} = {\cal L}^{(2)}+{\cal L}^{(4)}, 
\label{L}
\end{equation}
where $\cal L$ includes the derivative terms up to ${\cal O}(p^4)$.  The equation for $\theta$ is 
\begin{eqnarray}
&&f_\pi^2 \left({\theta^{\prime} \over \tau} + \theta^{\prime\prime} -\sin\theta \cos\theta |{\vec n}^{\prime}|^2 \right) + 16(\beta_1+\beta_2) \bigg[ \left({\theta^{\prime} \over \tau} + \theta^{\prime\prime}\right) \left({\theta^\prime}^2 + \sin^2\theta |{\vec n}^{\prime}|^2 \right)\nonumber\\ 
&&+ 2{\theta^\prime}^2 \theta^{\prime\prime} + 2 \sin^2\theta \, \theta^\prime
({\vec n}^{\prime} \cdot {\vec n}^{\prime\prime})+
\sin\theta \cos\theta |{\vec n}^{\prime}|^2 \left({\theta^\prime}^2 -
\sin^2\theta |{\vec n}^{\prime}|^2\right)\bigg] = 0.
\label{Op4th}
\end{eqnarray}
For $\vec n$, we get the following equation
\begin{eqnarray}
&&f_\pi^2 \left[2\cos\theta \, \theta^{\prime} \,{\vec n}^{\prime} + \sin\theta \left( {{\vec n}^{\prime} \over \tau} +  {\vec n}^{\prime\prime} + |{\vec n}^{\prime}|^2{\vec n}\right)\right] + 16(\beta_1+\beta_2)\Bigg\{\sin\theta \bigg[2\theta^{\prime}\theta^{\prime\prime}{\vec n}^{\prime}   \nonumber\\
&&+ {\theta^\prime}^2 \left({{\vec n}^{\prime} \over \tau}+{\vec n}^{\prime\prime}\right) + 
4\sin\theta \cos\theta \,\theta^{\prime} |{\vec n}^{\prime}|^2 {\vec n}^{\prime}
+\sin^2\theta |{\vec n}^{\prime}|^2 \left({{\vec n}^{\prime} \over \tau}+{\vec n}^{\prime\prime}\right) 
\nonumber\\
&&+2\sin^2\theta({\vec n}^{\prime} \cdot {\vec n}^{\prime\prime}){\vec n}^{\prime} +\left(\sin^2\theta |{\vec n}^{\prime}|^2 +  {\theta^\prime}^2\right) |{\vec n}^{\prime}|^2 {\vec n}\bigg] + 2\cos\theta\,{\theta^\prime}^3
{\vec n}^{\prime}
\Bigg\} = 0,
\label{Op4n}
\end{eqnarray} 
where we have used the constraint ${\vec n}\cdot{\vec n}^{\prime\prime} =
- |{\vec n}^{\prime}|^2$.

To solve the above coupled non-linear ordinary differential equations (\ref{Op4th}) and (\ref{Op4n}), we need to specify the boundary conditions.  In this work, we choose the boundary conditions for the ${\cal O}(p^4)$ fields $\theta$ and $\vec n$ and their derivatives to be the values of the corresponding ${\cal O}(p^2)$ solutions evaluated at a ``late'' proper time $\tau_l$, where the ${\cal O}(p^2)$ and ${\cal O}(p^4)$ solutions approximately coincide.  We choose $\tau_0 = 1/2m_\pi$, as a typical proper time where we expect the higher order interactions to become important, $\beta_1+\beta_2 = \beta = 2\times10^{-3}$, and $a=b=1$.  We pick $\tau_l = 10\tau_0$.  In this way, the solutions of the ${\cal O}(p^2)$ equations have definite values at $\tau_l$.  Numerically, we have $\tau_0 \approx 0.7$ fm, and $\tau_l \approx 7$ fm, at which we expect the ${\cal O}(p^2)$ solutions to approximate the ${\cal O}(p^4)$ solutions with good accuracy.

In Figs.(1) through (6), we present our numerical solutions for the fields $\theta$ and $\vec n$ and their derivatives.  Note that as a result of current conservation relations \cite{HS}, $n_1 = 0$ and $n_1^\prime = 0$, as in the 
case of the leading order solutions.  Each figure contains the ${\cal O}(p^2)$ solution, represented by the dashed line, and the ${\cal O}(p^4)$ solution, represented by the solid line.  The ${\cal O}(p^2)$ solutions are obtained numerically and agree with those of the analytic expressions in Eqs.(\ref{costh}) and (\ref{nb}).        

The solutions presented here are computed for $\tau \in [10^{-4}, 5/m_\pi]$ MeV$^{-1}$.  Figures (1), (2), and (3) show that the rapid oscillations of the ${\cal O}(p^2)$ solutions for $\theta$ and $\vec n$, near $\tau = 0$, no longer arise in the ${\cal O}(p^4)$ solutions, where the inclusion of the higher order terms seems to stabilize the solutions.  We tested the stabilization of the solutions
by comparing the values of the fields at proper times $\tau < 10^{-4}$ MeV$^{-1}$, and observing that whereas the leading order solutions oscillate more rapidly for smaller $\tau$, the values of the ${\cal O}(p^4)$ solutions do not oscillate or change significantly.

The ${\cal O}(p^4)$ and the ${\cal O}(p^2)$ field derivatives presented in Figs.(4), (5), and (6) show divergent behavior near $\tau = 0$.  The analytic
${\cal O}(p^2)$ solutions of Eqs.(\ref{costh}) and (\ref{nb}) yield divergent and oscillatory derivatives for $\tau = 0$.  To check the divergence of the ${\cal O}(p^4)$ derivatives, we examined their behavior for $\tau < 10^{-4}$ MeV$^{-1}$.  We found that the magnitudes of the derivatives did not stabilize and continued to grow with decreasing $\tau$.

In order to understand the behavior of the ${\cal O}(p^4)$ solutions mentioned above, we examine Eq.(\ref{Op4th}) in the limit $\tau \to 0$, for the case of
constant $\vec n$.  The solutions with constant $\vec n$ are related to those with spacetime dependence by chiral rotations \cite{HS}.  In this case, Eq.(\ref{Op4th}) reduces to
\begin{equation}
f_\pi^2 \left({\theta^{\prime} \over \tau} + \theta^{\prime\prime}\right)+ 16(\beta_1+\beta_2)\left({\theta^{\prime} \over \tau} + 3\theta^{\prime\prime}\right){\theta^\prime}^2 = 0.
\label{Op4thr}
\end{equation}
Let us assume that for small $\tau$, the behavior of the field $\theta$ is given by
\begin{equation}
\theta = \tilde \theta + \left({\tau \over \tilde \tau}\right)^p,
\label{stau}
\end{equation}
where $\tilde \theta$ and $\tilde \tau$ are constants.  We expect to find a solution for $\theta$ with $0<p<1$ that tends to a constant $\tilde \theta$ but has a divergent derivative, as $\tau \to 0$.  Upon substituting the expression in Eq.(\ref{stau}) for $\theta$ into Eq.(\ref{Op4thr}), we get the following equation
\begin{equation}
f_\pi^2[p + p(p-1)] + 16(\beta_1+\beta_2)
[p + 3p(p-1)] p^2 \left({\tau \over \tilde \tau}\right)^{2p}
\left({1 \over \tau}\right)^2 = 0.
\label{peq}
\end{equation}
In the absence of the ${\cal O}(p^4)$ terms, we only have the terms proportional to $f_\pi^2$.  In this case, we get $p = 0$, which is consistent with the  solution $\theta \sim \ln (\tau/\tilde \tau)$  of the equation $\theta^{\prime}/\tau + \theta^{\prime\prime} =0$.  However, we see that for small $\tau$, the terms proportional to $\beta_1+\beta_2$ will be dominant if $0<p<1$.  Thus, for $\tau \to 0$, we ignore the terms proportional to $f_\pi^2$.  We then get
\begin{equation}
p={2 \over 3},
\label{p}
\end{equation}
in agreement with our prior assumption that $0<p<1$.

The numerical solutions presented here are obtained in the context of a systematic derivative expansion.  Thus, it is the magnitudes of the derivatives that establish the region of validity of the solutions.  We take the maximum magnitude of the derivative of a field below which the expansion is valid to be  $p_{max} \sim 500$ MeV.  The graphs in Figs.(4), (5), and (6) show that the magnitudes of the ${\cal O}(p^4)$ field derivatives stay below $p_{max}$ for values of $\tau$ down to $\tau = 10^{-3}$ MeV$^{-1} \sim \Lambda_\chi^{-1}$, which is as small a proper time as we can consider in our formalism.  In contrast, the magnitudes of the ${\cal O}(p^2)$ field derivatives begin to exceed $p_{max}$ at proper times $\tau \lsim (2-4)\times 10^{-3}$ MeV$^{-1}$, indicating a smaller range of validity for these solutions.  Hence, our results suggest that the ${\cal O}(p^4)$ solutions can be used to study the qualitative evolution of the DCC down to a length scale of $\sim 0.2$ fm, but the ${\cal O}(p^2)$ solutions lose their qualitative validity below a length scale of $\sim (0.5-0.8)$ fm.

\section{Conclusions}

We derived the ${\cal O}(p^4)$ equations of motion for the DCC produced in an idealized relativistic collision, using the non-linear sigma model Lagrangian without the mass terms.  We presented our numerical solutions for the ${\cal O}(p^4)$ equations of motion.  These higher order corrections to the ${\cal O}(p^2)$ solutions are only important for small values of proper time $\tau \ll 1/m_\pi$.  Hence, the absence of the mass terms, important only for $\tau \grtsim 1/m_\pi$, does not introduce significant qualitative changes in the early proper time solutions.  Our ${\cal O}(p^4)$ solutions for the fields $\theta$ and $\vec n$, parametrizing the DCC field configuration, stabilize for small values of proper time, whereas the ${\cal O}(p^2)$ solutions oscillate rapidly with decreasing $\tau$.  We presented a qualitative explanation for the behavior of the solutions by studying the equations of motion, in a special case and in the limit where $\tau \to 0$.  A measure of the validity of our derivative expansion is the magnitude of the field derivatives.  We took $p_{max} \sim 500$ MeV to be the maximum value for the magnitude of a field derivative beyond which the solutions cannot be trusted.  Using this criterion, the ${\cal O}(p^4)$ solutions were deemed reliable down to $\tau \sim 0.2$ fm, whereas the ${\cal O}(p^2)$ solutions were considered no longer qualitatively valid below $\tau \sim (0.5 - 0.8)$ fm.  We take our solutions to represent a qualitative measure of the behavior of the DCC.  Our results suggest that this qualitative behavior can be studied within the non-linear sigma model down to a length scale of $\sim 0.2$ fm, once the ${\cal O}(p^4)$ derivative corrections to the Lagrangian are included.  A length scale of $\sim 0.2$ fm corresponds to the energy scale $\Lambda_{\chi} \sim 1$ GeV, where chiral symmetry is restored and the non-linear sigma model formalism is no longer valid.  Hence, we do not expect the inclusion of the higher order terms beyond ${\cal O}(p^4)$ in the chiral Lagrangian to enable us to study even earlier proper times in the evolution of the DCC.               
 
\section*{Acknowledgements}
I would like to thank Krishna Rajagopal who generously offered his time for many helpful conversations.  It is also a pleasure to thank Mark Wise, for his advice on a number of issues that were discussed in this paper, and Zoltan Ligeti, for his help with the diagrams.  Peter Cho, Martin Gremm, Adam Leibovich, Iain Stewart, and Eric Westphal are also thanked for various discussions.  This work was supported in part by the U.S. Dept. of Energy under Grant No. DE-FG03-92-ER40701.

\begin{figure}[[htbp]    
\centerline{\epsfxsize=8truecm \epsfbox{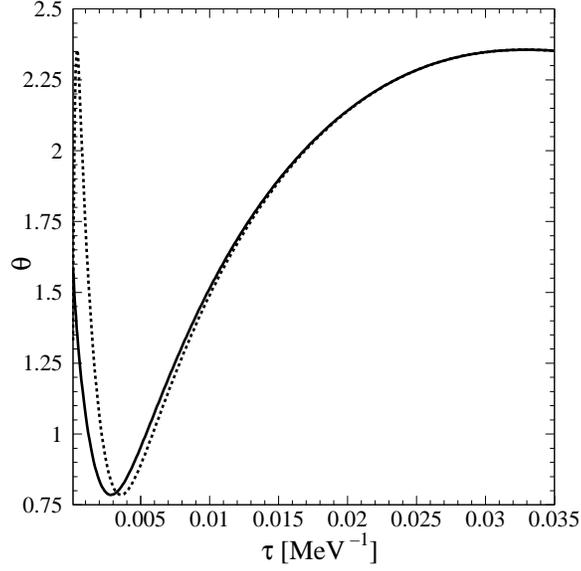}}
\caption[1]{The dashed and the solid lines represent the ${\cal O}(p^2)$ and the ${\cal O}(p^4)$ solutions for $\theta$, respectively.  The ${\cal O}(p^2)$ solution for $\theta$ oscillates increasingly rapidly with decreasing $\tau$, as Eq.(\ref{costh}) implies, while the ${\cal O}(p^4)$ solution does not oscillate as $\tau \to 0$.  Similar comments apply to the solutions for $n_2$ and $n_3$ in Figs. (2) and (3), respectively.}
\end{figure}

\begin{figure}[[htbp]    
\centerline{\epsfxsize=8truecm \epsfbox{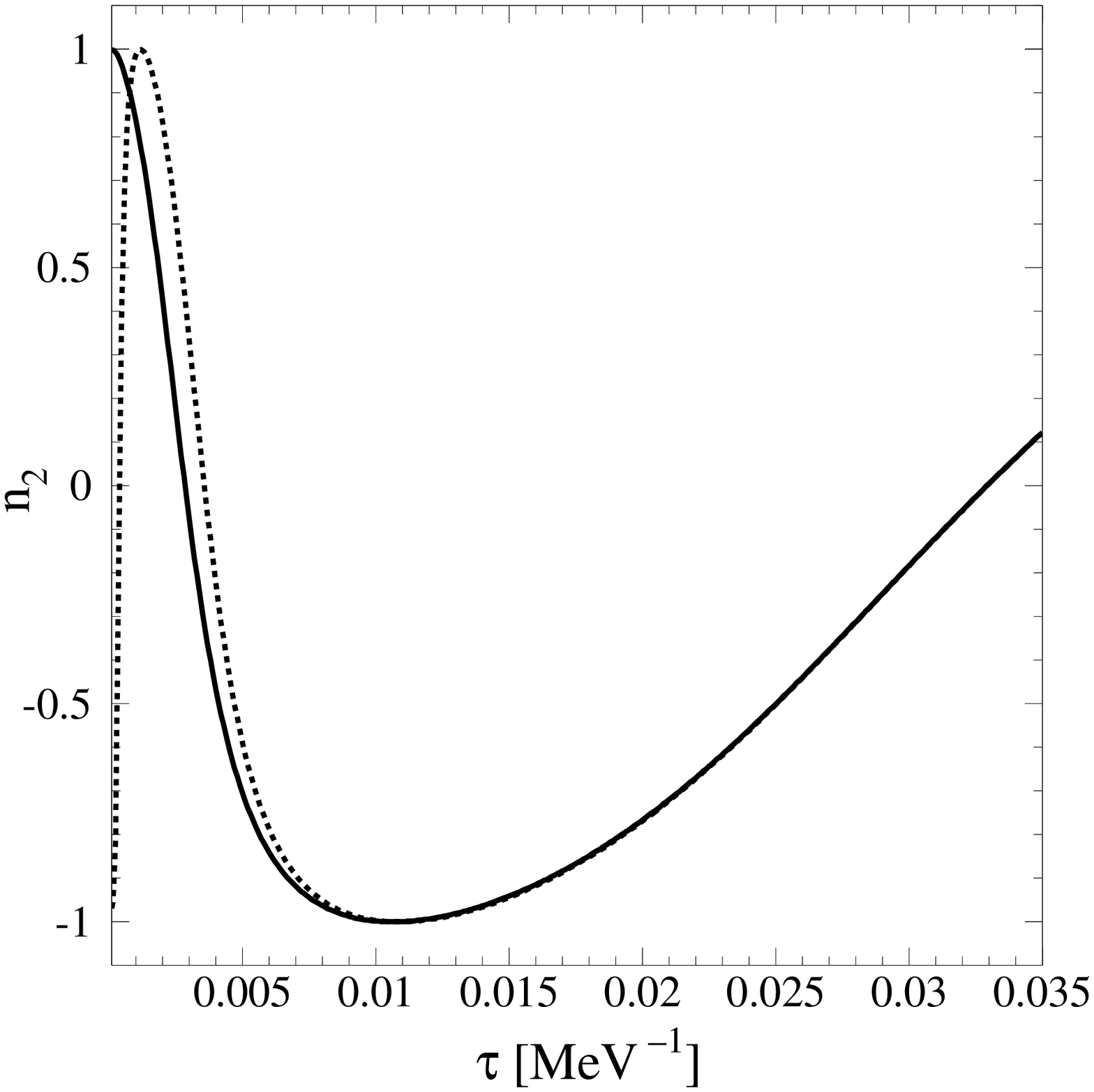}}
\caption[2]{The dashed and the solid lines represent the ${\cal O}(p^2)$ and the ${\cal O}(p^4)$ solutions for $n_2$, respectively.}
\end{figure}

\begin{figure}[[htbp]    
\centerline{\epsfxsize=8truecm \epsfbox{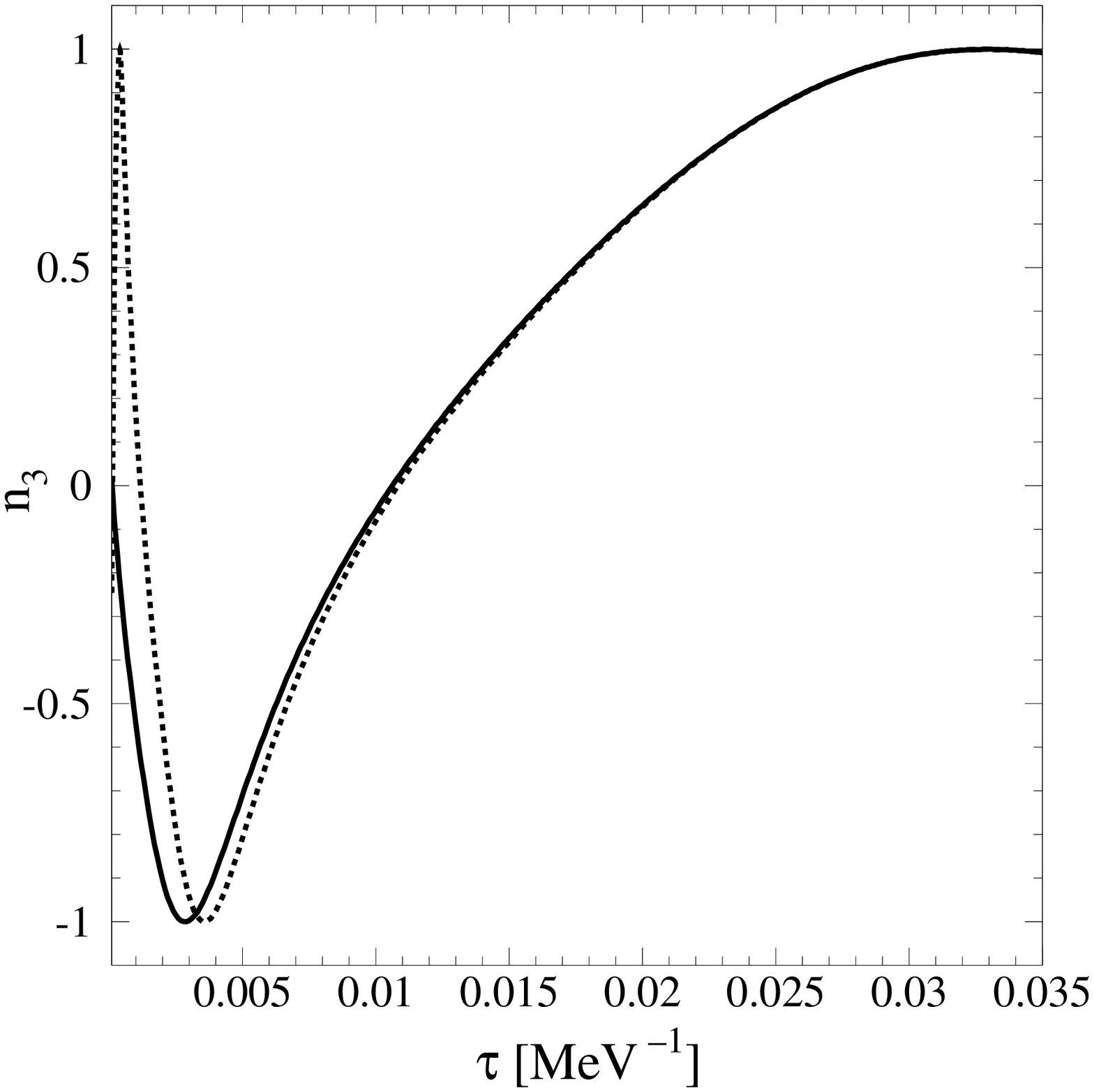}}
\caption[3]{The dashed and the solid lines represent the ${\cal O}(p^2)$ and the ${\cal O}(p^4)$ solutions for $n_3$, respectively.}
\end{figure}

\begin{figure}[[htbp]    
\centerline{\epsfxsize=8truecm \epsfbox{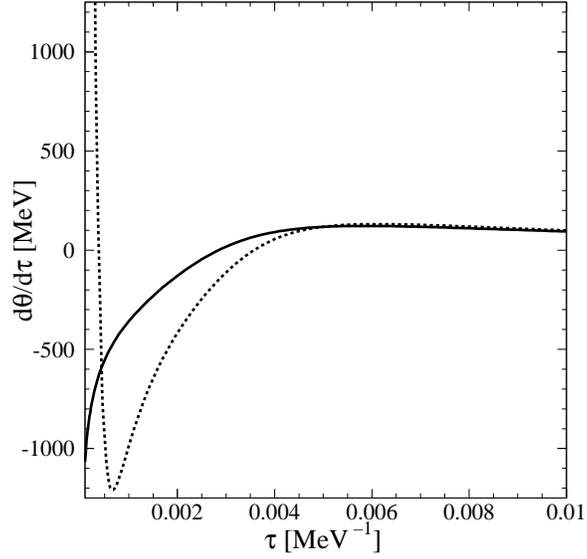}}
\caption[4]{The dashed and the solid lines represent the ${\cal O}(p^2)$ and the ${\cal O}(p^4)$ solutions for $\theta^\prime$, respectively.  The 
${\cal O}(p^2)$ solution for $\theta^\prime$ oscillates and has a divergent magnitude, while the ${\cal O}(p^4)$ solution for $\theta^\prime$ diverges in magnitude, but does not oscillate.  Similar comments apply to the solutions for $n_2^\prime$ and $n_3^\prime$ in Figs. (5) and (6), respectively.}
\end{figure}

\begin{figure}[[htbp]    
\centerline{\epsfxsize=8truecm \epsfbox{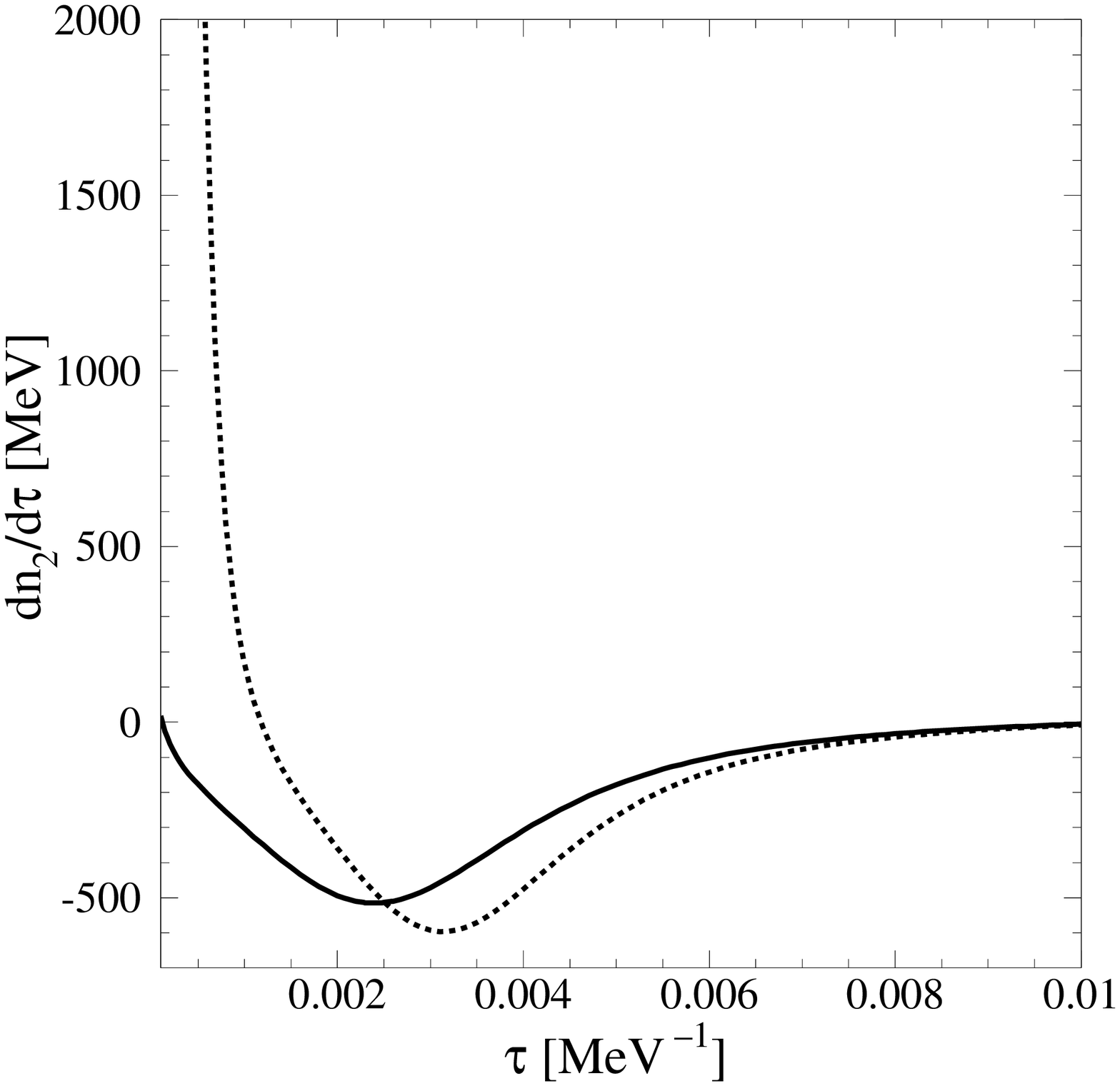}}
\caption[5]{The dashed and the solid lines represent the ${\cal O}(p^2)$ and the ${\cal O}(p^4)$ solutions for $n_2^\prime$, respectively.}
\end{figure}

\begin{figure}[[htbp]    
\centerline{\epsfxsize=8truecm \epsfbox{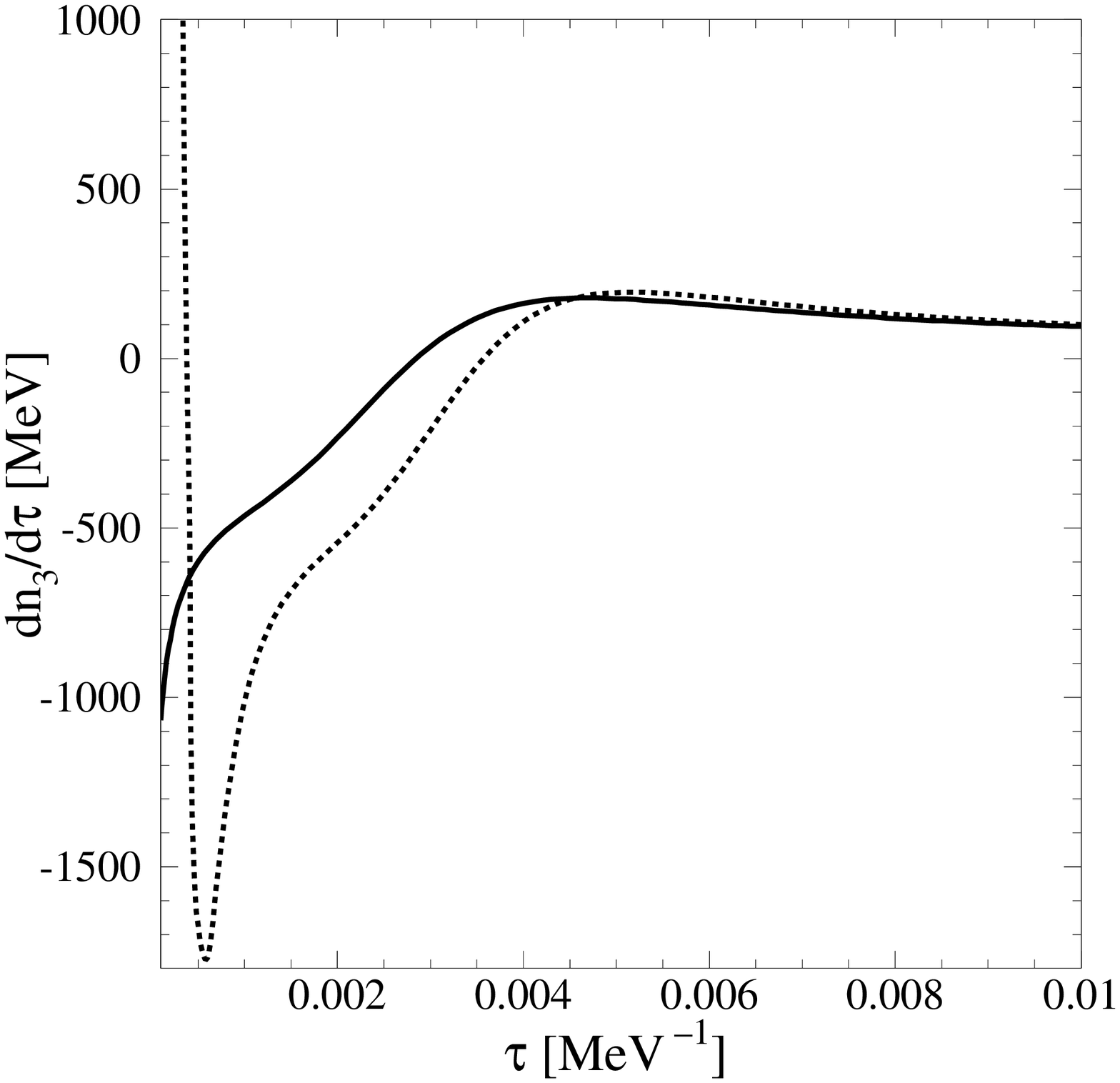}}
\caption[6]{The dashed and the solid lines represent the ${\cal O}(p^2)$ and the ${\cal O}(p^4)$ solutions for $n_3^\prime$, respectively.}
\end{figure}

\end{document}